\documentclass[prd,aps,showpacs,showkeys]{revtex4-1}
\usepackage{bm,amsfonts,latexsym,amsmath,amssymb,amsbsy,graphicx}

\newcommand{\bb}{\begin{eqnarray}}
\newcommand{\ee}{\end{eqnarray}}
\newcommand{\ba}{\begin{align}}
\newcommand{\ea}{\end{align}}

\begin{document}

\title{\bf Plane density of induced vacuum charge in a supercritical Coulomb potential}
\author{V.R. Khalilov \footnote{Corresponding author}}\email{khalilov@phys.msu.ru}
\affiliation{Faculty of Physics, M.V. Lomonosov Moscow State University, 119991,
Moscow, Russia}
\author{I.V. Mamsurov}
\affiliation{Faculty of Physics, M.V. Lomonosov Moscow State University, 119991,
Moscow, Russia}

\begin{abstract}
An expression for the density of a planar induced vacuum charge is obtained
in a strong Coulomb potential in coordinate space. Treatment is based on a self-adjoint
extension approach for constructing of the Green's function of a charged fermion in this potential.
Induced vacuum  charge density is calculated and analyzed at the subcritical
and supercritical Coulomb potentials for massless
and massive fermions.  The behavior of the obtained vacuum charge density is investigated
at long and short distances from the Coulomb center. The  induced vacuum charge has a  screening sign. Screening of a Coulomb impurity in graphene is briefly discussed. We calculate the real vacuum polarization charge density that acquires the quantum electrodynamics vacuum in the supercritical
Coulomb potential due to the so-called real vacuum polarization. It is shown that
the vacuum charge densities essentially  differ in massive and  massless cases.
We expect  that our results can, as a matter of principle, be tested
in graphene with a supercritical Coulomb impurity.
\end{abstract}



\maketitle

\section{Introduction}

%PACS numbers: 73.22.Pr, 81.05.ue, 03.65.Pm

The vacuum of the quantum electrodynamics  and
the induced vacuum polarization in a strong Coulomb field produced
by a heavy atomic nucleus  have been studied a long time \cite{003,004,grrein,01,02,03,04}.
  When the nuclear charge $Z|e|$ ($e$ is the electron charge) is increased from subcritical
to supercritical values then the lowest electron energy level (in the regularized Coulomb potential) dives into the negative energy continuum and becomes a resonance with complex ``energy'' $E$ signaling  the instability of the quantum electrodynamics vacuum in the supercritical range. The  nuclear  charge $Z_{cr}|e|$ for which the lowest energy level descends to the negative-energy continuum boundary $-m$ is called the critical charge for the ground state. The critical charge is obviously related to the fine structure constant $1/137$ and  the number  $Z_{cr}\sim 170$ \cite{05}. It has been understood that vacuum polarization
effects predict wonderful phenomenon such as electron-positron pair production from vacuum.
 This fundamental phenomenon due to the instability
of the quantum electrodynamics vacuum in the supercritical Coulomb potential
is difficult to probe experimentally and is unlikely to be observed in foreseen future.

However, similar phenomena are likely to be revealed in graphene with charged
impurities because the corresponding ``effective fine structure constant"
is large $\sim 2$ and a cluster of charged impurities  can produce the supercritical
Coulomb potential. Thus, it is to be expected that the phenomenon
such as the electron-hole pair production is now within experimental
reach in a graphene (see, \cite{gr0,gr1,gr2}).
 In graphene, the electrons near the Fermi surface can be described
in terms of an effective Lorentz-invariant theory with their energy
determined by Dirac's dispersion law for massless fermions \cite{6,7,review},
which allows to consider graphene as the condensed matter analog of the quantum electrodynamics
in 2+1 dimensions \cite{datdc,ggvo}. The massless case turn out to be rather more complicated
as compared massive one since an infinite number of quasi-stationary states (resonances) emerges
in the ``hole" sector in the presence of a supercritical Coulomb potential \cite{vp11a,as11b,ggg,20}.

Vacuum polarization of graphene  with a Coulomb impurity was studied  in  \cite{7,review,vp11a,as11b,13a,as112,bsaso,kn11c,12,vmpvnk,yn1}.
The vacuum polarization of the massive charged fermions can also be of interest
for graphene with Coulomb impurity \cite{13}. For massive fermions the vacuum polarization
charge density  behaves differently from the massless ones.

Here we study the density of a planar induced vacuum charge in a strong  Coulomb
potential.   The problem is considered by means of a self-adjoint extension approach,
recently used by the authors for the vacuum polarization problem
of massless charged fermions in Aharonov--Bohm potential (\cite{kh2}) as well as in the superposition of  Coulomb and Aharonov--Bohm  potentials  \cite{khmam}.    We express the density of
an induced charge in the vacuum via the exact Green function, constructed
from solutions of the self-adjoint two-dimensional Dirac Hamiltonians
with a strong Coulomb potential. The self-adjoint Dirac Hamiltonians are not unique
and can be specified by a self-adjoint extension parameter which implies
 additional nontrivial boundary conditions on the wave functions at the origin \cite{17}.
Physically, the self-adjoint extension parameter can be interpreted, for example, in terms of
the radius $R$ of a real nucleus (or a Coulomb impurity) that generates
a cut (at distances $R$) Coulomb potential. It is well to note that the self-adjoint extension
approach was used  for various problems in the  Aharonov--Bohm-like fields
in \cite{phg,as0,sil}.

We also address the pure (vector) Coulomb problem interacting with a
scalar potential $U(r)=-b/r, b>0$ located at the origin and  argue that the ground fermion state in
the vector Coulomb potential is stabilized in the presence of a scalar potential.
It is useful to remind that the Dirac Hamiltonian with a vector potential
does not exhibit a charge conjugation symmetry because a charge coupling
treats particles and antiparticles differently while the Dirac Hamiltonian
a scalar potential is added to the mass term of the Dirac Hamiltonian
and, therefore, a scalar coupling treats  particles and antiparticles
similarly. This coupling has been used to consider  various physical
problems, for instance, in \cite{hokh2,aal,erfm,wcasc,lses}.

We shall adopt the units where $c=\hbar=1$.

\section{Green's function for the self-adjoint two-dimensional Dirac Hamiltonians}

The Dirac Hamiltonian for a fermion of the mass $m$ and charge
$e=-e_0<0$, which contains a parameter $s=\pm 1$ to label two types of fermions \cite{27} or
to characterize the fermion spin ("up" and "down") \cite{4} in vector
($A_0(r) =Ze_0/r\equiv a/e_0r$, $A_r=0, A_{\varphi}=0, a>0$) and scalar ($U(r)=-b/r, b>0$)
Coulomb potentials  is
\bb
 H_D=\sigma_1P_2-s\sigma_2P_1+\sigma_3 [m+U(r)]-e_0A_0(r),\label{diham}
\ee
where $P_\mu = -i\partial_{\mu} - eA_{\mu}$ and $\gamma^{\mu}$ is represented in terms of the  two-dimensional Pauli matrices
$\gamma^0= \sigma_3,\quad \gamma^1=is\sigma_1,\quad \gamma^2=i\sigma_2$.
The  total angular momentum operator $J=-i\partial/\partial\varphi+ s\sigma_3/2$ commutes with  $H_D$. Eigenfunctions of the Hamiltonian (\ref{diham}) are (see, \cite{khho,khpr,khho0})
\bb
 \Psi(t,{\bf r}) = \frac{1}{\sqrt{2\pi r}}
\left( \begin{array}{c}
f(r)\\
g(r)e^{is\varphi}
\end{array}\right)\exp(-iEt+il\varphi)~, \label{three}
\ee
where $r=\sqrt{x^2+y^2}, \varphi=\arctan(y/x)$ are  polar coordinates, $E$ is  the fermion energy, $l$ is an integer.
The wave function $\Psi$ is an eigenfunction of the
operator $J$ with eigenvalue $j=\pm (l+s/2)$ in terms of the angular momentum $l$ and
\bb \check h F(r)= EF(r), \quad F(r)=\left(
\begin{array}{c}
f(r)\\
g(r)\end{array}\right), \label{radh}\ee
where
\bb
\check h=is\sigma_2\frac{d}{dr}+\sigma_1\frac{l+s/2}{r}+\sigma_3(m-\frac{b}{r})-\frac{a}{r}. \label{radh0}
\ee

The planar vacuum  current density $j_{\mu}({\bf r})$  can be expressed
via the Green's function  as
\bb
 j_{\mu}({\bf r})=-\frac{e}{2}\int\limits_{C}\frac{dE}{2\pi i}{\rm tr}G({\bf r}, {\bf r'}; E)|_{{\bf r}={\bf r'}}\gamma_{\mu},
\label{cur0}
\ee
where $C$ is the integration path along the real axis $E$ in the complex plane of $E$.
Main role in this expression plays the radial partial Green's function $G_l(r, r'; E)$ that
must satisfy appropriate boundary conditions at $r\to \infty$ and $r\to 0$ with $r'$ fixed.
Then, the radial Green's function can be constructed by means of the regular and irregular
solutions of the radial Dirac equation $(\check h-E)U(r)=0$ as follows (see, also \cite{grrein})
\bb
G_l(r, r'; E)\gamma^0=\frac{1}{{\rm W}(E)}[\Theta(r'-r)U_R(r)U^{\dagger}_I(r')+
\Theta(r-r')U_I(r)U^{\dagger}_R(r')].
\label{green5}
\ee
Here ${\rm W}(E)$ is the Wronskian and the regular solution $U_R(r)$ is integrable near $r\to 0$,
while the irregular solution $U_I(r)$  is integrable at $r\to\infty$.

The Hamiltonian (\ref{radh0}) is singular and requires the supplementary definition to be treated as a self-adjoint quantum-mechanical operator. The additional specification of its domain can be
given by means of the (real) self-adjoint extension parameter $\xi$ in terms
of  boundary conditions at the origin for any solution $F(r)$  \cite{17,khlee0,khlee}.
\bb
 (F^{\dagger}(r)i\sigma_2 F(r))|_{r=0}= (\bar f_1f_2-\bar f_2f_1)|_{r=0} =0, \label{bouncon}
\ee
which shows that  the probability current density is equal to zero at the origin.

The regular and irregular solutions of Eq. (\ref{radh0}) can expressed
via the Whittaker functions $M_{c,d}(x)$ and $W_{c,d}(x)$ as
\bb
F_R = \left( \begin{array}{c} f_R(r, \gamma, E) \\ g_R(r, \gamma, E) \end{array} \right),
F_I = \left( \begin{array}{c} f_I(r, \gamma, E) \\ g_I(r, \gamma, E) \end{array} \right),
\label{WF1}
\ee
where
\bb
f_R(r, \gamma, E)=\frac{\sqrt{m+E}}{x} \left( A_R M_{(aE+mb)/\lambda +s/2, \gamma}
(x) + C_R M_{(aE+mb)/\lambda -s/2, \gamma} (x) \right), \phantom{mmmmmmmm}\nonumber \\
g_R(r, \gamma, E)=\frac{\sqrt{m-E}}{x} \left( A_R M_{(aE+mb)/\lambda +s/2, \gamma}
(x) - C_R M_{(aE+mb)/\lambda -s/2, \gamma} (x) \right), \phantom{mmmmmmmm}\nonumber \\
\frac{C_R}{A_R}=\frac{s\gamma -(aE+mb)/\lambda}{\nu+(ma+Eb)/\lambda},\phantom{mmmmmmmmmmmmmmmmmmmmm}
\label{base}
\ee
\bb
f_I(r, \gamma, E)=\frac{\sqrt{m+E}}{x} \left( A_I W_{(aE+mb)/\lambda +s/2, \gamma}
(x) + C_I W_{(aE+mb)/\lambda -s/2, \gamma} (x) \right),\phantom{mmmmmmmm} \nonumber \\
g_I(r, \gamma, E)=\frac{\sqrt{m-E}}{x} \left( A_I W_{(aE+mb)/\lambda +s/2, \gamma}
(x) - C_I W_{(aE+mb)/\lambda -s/2, \gamma} (x) \right), \phantom{mmmmmmmm}\nonumber \\
\frac{C_I}{A_I}=[(ma+Eb)/\lambda-s\nu]^s. \phantom{mmmmmmmmmmmmmmmmmm}
\label{BASE}
\ee
Here
\bb
x=2\lambda r, \quad \lambda = \sqrt{m^2 - E^2}, \quad \gamma = \sqrt{\nu^2
-a^2+b^2}, \quad \nu=|l+s/2|, \label{Note}
\ee
$A_R, A_I, C_R, C_I$ are numerical coefficients and
we take into account that the asymptotic behavior of the functions  $M_{c,d}(x)$, $W_{c,d}(x)$
as $x\to 0$ is given by $M_{c,d}(x)\sim x^{d+1/2}$, $W_{c,d}(x)\sim x^{-d+1/2}$ and that
$W_{c,d}(x)\sim e^{-x/2} x^c$ as $x\to \infty$.
All the fermion states are doubly degenerate with respect
to the spin parameter $s$.
We set $\sqrt{\nu^2-a^2+b^2}\equiv \gamma$ for $\nu^2\geq a^2-b^2$
and $i\sqrt{a^2-b^2-\nu^2}\equiv i\sigma$ for $a^2-b^2>\nu^2$ and call these regions
subcritical and supercritical ones, respectively. In subcritical region,
defining the energy spectra by standard quantum mechanical methods encounters no problems.
Relevant quantum system in the lowest state (with $l=0$) becomes  unstable in the supercritical region
for $\sqrt{a^2 - b^2}>1/2$, thus  scalar potential stabilizes the system.
It should also be emphasized that  the system never occurs in the supercritical region in the presence of scalar potential with coupling $b>\sqrt{a^2-1/4}$.

In the subcritical range, only solutions $F_R(r)$ vanishing at $r=0$ are the regular ones for $\gamma\geq 1/2$  while  the linear superposition $U_R(r)$  \cite{17,khlee}
\bb
U_R(r)=F_R(r)+\xi F_I(r)
\label{mainf}
\ee
should be chosen as the regular ones for $1/2>\gamma>0$; $U_R(r)$ satisfies the self-adjoint boundary
condition (\ref{bouncon}).
Nevertheless, one can show that the  contribution into the induced charge density  is very small
for $1/2>\sqrt{(l+s/2)^2-a^2+b^2}>0$ (compared with the contribution for $\sqrt{(l+s/2)^2-a^2+b^2}\geq 1/2$)  in the subcritical range at any $\xi$;  therefore, one can put $\xi=0$ but choose as the regular solutions  the functions $F_R(r)$ for all $\gamma>0$ taking into account the small contribution from the range $1/2>\gamma>0$ in this way.
  Thus, in the subcritical range the Green's function is completely determined:
\bb
{\rm tr} G_{\nu} ({\bf r}, {\bf r'}; E)|_{{\bf r}={\bf r'}} \gamma^0 = \sum_{s={\pm 1}}
\sum^{\infty}_{l=-\infty}\frac{f_I f_R + g_I g_R}{2\pi s{\rm W}(E, \gamma)},
\label{TRACE}
\ee
where
\bb
{\rm W}(E, \gamma) = (g_R f_I - f_R g_I) = -2 A_R A_I \frac{\Gamma
(2\gamma)}{\Gamma (\gamma + 1/2 -s/2 -(aE+mb)/\lambda)}
\frac{s\gamma}{\nu+(ma+Eb)/\lambda}
\label{wr1}
\ee
and $\Gamma(z)$ is the Gamma function \cite{GR}.

 Generally speaking the self-adjoint parameter is related to
the behavior of the upper (lower) component of solutions
(\ref{WF1})   at the origin.
Particularly, the case $\xi=0$ ($\xi=\infty, -\infty\thicksim{\infty}$)  is equivalent to insisting that the upper (lower) component stays regular at the origin. If $\xi\neq 0, \infty$ both components of the doublet contain singular terms at the origin.

In the supercritical regime, $\gamma=i\sigma$, the above two solutions $F_R(r)$ and $F_I(r)$ become oscillatory  at $r\to 0$ with the imaginary exponent.  Both solutions
are now equally important. So as the regular solutions $U_R(r)$ have to be chosen
their linear superposition. Therefore,
the time component of induced charge (electron) density (\ref{cur0}) can be represented as follows
\bb
j_0({\bf r})=j_{sub}({\bf r})+j_{super}({\bf r}),
\label{cursum}
\ee
where $j_{sub}({\bf r})$ ($j_{super}({\bf r})$) contributes to $j_0({\bf r})$ from the subcritical  (supercritical) range and these terms have to be treated separately.
One can easily understand that only the case $a>b$ is of interest,
and we hence assume that $b=0$ in what follows, without restricting the generality.

First we calculate $j_{sub}({\bf r})$.
Summing over $s$ in (\ref{TRACE}), we obtain
\bb
{\rm tr} G_{\nu} ({\bf r}, {\bf r'}; E)|_{{\bf r}={\bf r'}} \gamma^0 = -\frac{1}{2\pi \lambda^2 r^2}
\sum^{\infty}_{l=-\infty} \frac{\Gamma(\gamma -
aE/\lambda)}{\Gamma(2\gamma +1)} \left[ (m^2 a/\lambda +E(x
-2aE/\lambda -1 )) M_{aE/\lambda +1/2, \gamma} (x) W_{aE/\lambda
+1/2, \gamma} (x) + \right. \nonumber \\
\left. + m^2a [(\gamma -aE/\lambda)/\lambda] M_{aE/\lambda -1/2,
\gamma} (x) W_{aE/\lambda -1/2, \gamma} (x) +
Ex\frac{d}{dx}(M_{aE/\lambda +1/2, \gamma} (x) W_{aE/\lambda +1/2,
\gamma} (x)) \right]. \phantom{mmmm}\label{Calcul1}
\ee
Here and below  $\nu =l+1/2$ and $\gamma = \sqrt{(l+1/2)^2-a^2}$.

%In the subcritical range  the singularities of $G_{\nu}({\bf r}, {\bf r'}; E)$ in $E$ are
%simple poles for $-m<E<m$, and two cuts ($(-\infty,-m]$ and $[m,\infty)$) for $|E|\geq m$ \cite{20}.
%  The path $C$ runs along the real $E$ axis as follows:
%$C=C_-+C_p+C_+$, where $C_-$ is the path along the negative
%real $E$ axis (${\rm Re}E<0$) from $-\infty$ to $0$ turning
%around at $E=0$ with positive
%orientation, $C_p$ is a circle around the bound states singularities
%with $-m< E<0$ (if we chose $E_F=-m$),
%and $C_+$ is the path along the positive real $E$ axis (${\rm Re}E>0$)
%from $\infty$ to $0$ but with negative orientation (i.e. clockwise path)
%turning around at $E=0$ \cite{grrein,kh1}.

It is convenient to deform the integration path $C$ on the imaginary  $E$ axis (see, \cite{20,kh2}):
\bb
j_{sub}({\bf r}) = -e \int\limits_{-\infty}^{\infty}\frac{dE}{2\pi}{\rm tr}
G_{\nu} ({\bf r},{\bf r}, iE) \gamma^0. \label{Density}
\ee
By means of formula \cite{GR}
\bb
M_{aE/\lambda \pm 1/2,\gamma}(x) W_{aE/\lambda \pm 1/2,\gamma}(x)
=\frac{x\Gamma(2\gamma +1)}{\Gamma(1/2 +\gamma -aE/\lambda \mp 1/2)}
\int\limits_{0}^{\infty} e^{-x\cosh s}
[\coth(s/2)]^{2aE/\lambda \pm 1} I_{2\gamma}(x\sinh s) ds, \phantom{mmm}
\label{Form}
\ee
we rewrite the induced charge density in the form
\bb
j_{sub}(r) = - \frac{8e}{\pi^2 r } \sum_{l=0}^{\infty}
\int\limits_{0}^{\infty} dE \int\limits_{0}^{\infty} dt
e^{-2\lambda r \coth t} \left( a\cos(2aE/\lambda) \coth t I_{2\gamma}
(2\lambda r/\sinh t)  -\right. \nonumber \\
- \left. \frac{Er}{\sinh t} \sin(2aE/\lambda)
I^{\prime}_{2\gamma} (2\lambda r/\sinh t) \right).
\label{Res1}
\ee
where  $\lambda =\sqrt{m^2+E^2}$, $I_{\mu}(z)$ is the modified Bessel function of the first kind  and the prime (here and below) denotes the derivative of function with respect to argument.
 We note that $j_{sub}(r)$ is odd with respect to  $a$.

\section{Renormalized induced charge}

Since the presence of external fields do not give rise to additional divergences
in expressions of perturbation theory it is enough  to carry out the renormalization  in the subcritical range. We note that the expansion (\ref{Res1}) of $j_{sub}(r)$ in terms of $a$
contains only odd powers of this parameter. Expression (\ref{Res1}) calls
for renormalization, which can be  carried out on the basis of
the obvious physical requirement of vanishing of the total induced charge.
This can made because the induced charge density diminishes rapidly at distances $r\gg 1/m$.
The renormalization can be performed as well as in the conventional quantum
electrodynamics in momentum space:
\bb
j_{sub}(z)\equiv \rho(z) =\int d{\bf r} e^{i{\bf p\cdot r}} j_{sub} (r)
=\frac{e}{\pi} \sum_{l=0}^{\infty} \int\limits^{\infty}_{0} dx
\int\limits^{\infty}_{0} dt \int\limits^{\infty}_{0} dy \frac{\sinh
t}{\sqrt{1+x^2}} e^{-y \cosh t} J_{0} (zy \sinh t/2\sqrt{1+x^2}) g(y,t), \nonumber \\
g(y,t)=2\frac{xy}{\sqrt{1+x^2}}I^{\prime}_{2\gamma}(y)\sin(ct)
- 4a I_{2\gamma} (y) \coth t \cos(ct). \phantom{mmmmmmmmmmmmmmm}\label{Imp}
\ee
Here  $z=|{\bf p}|/m\equiv p/m$, $x=E/m$, $y=2mr\sqrt{1+x^2}/\sinh t$, $c=2ax/\sqrt{1+x^2}$.

Let us define the renormalized induced charge  as $\rho_r(z)=\lim_{\Lambda\to \infty}[\rho(z)-\lim_{z\to 0}\rho(z)]$ introducing
a finite upper limit of integration for $|E|<\Lambda$.
%Obviously the  function
%$\rho_r(z)$ can depend only on the dimensionless ratio $p/m$.
As $a<1/2$, the terms of different order in $a$ behave differently.
We can see it in terms of perturbation theory. Indeed, the
linear in $a$ term  corresponds to the diagram of the polarization operator
in the one-loop approximation  and its renormalization  coincides with
the usual procedure of renormalizing the polarization operator.
The terms proportional to $a^3$ correspond to
diagrams of the type of photon scattering by photon and, in difference on the
case of the 3D quantum electrodynamics (see \cite{01,02,03,ms0}) they are finite. However their
regularization must still be carried out in the considered case  due to the requirements of gauge  invariance, which, in particular,  determine the behaviors of the scattering amplitude at small  $p/m$.

{\bf Massless case}.
We shall first consider  the more complicated case with $m=0$.
The leading term of the asymptotics of the function $\rho_r(z)$ at $m\to 0$ is a constant,
 $q_{ind}$.  Hence, the induced charge density in coordinate space can be represented as
 \bb \rho_r({\bf r})=q_{ind}\delta({\bf r})+\rho_{dist}({\bf r}). \label{qtot} \ee
The induced charge $q_{ind}$ is negative (see below), the distributed
charge density $\rho_{dist}({\bf r})$  is positive and the total distributed charge
is $-q_{ind}$.

For the renormalized induced  charge in the subcritical region $q_{ind}$ we obtain (see Appendix)
\bb
q_{ind} = q_1(e_0a) + q_r(e_0a). \label{Tot}
\ee
Here
\bb q_1(e_0a)= \frac{2ea}{\pi}
\sum_{l=0}^{\infty}\left( 2(l+1/2) \psi^{\prime}(l+1/2) - 2 - \frac{1}{l+1/2}
\right)=-\frac{e_0a\pi}{4}
\label{Q1}
\ee
contains the terms of order  $a$,
\bb
q_r(e_0a) = -\frac{2e_0}{\pi} \sum^{\infty}_{l=0}{\rm Im} \left[ \ln (\gamma - ia)(\Gamma
(\gamma - ia))^2 - \phantom{mmmmmm} \right.\nonumber\\  \left. -2(\gamma -ia) \psi (\gamma -ia) +  \frac{ia}{l+1/2} -
 2ia (l+1/2) \psi^{\prime} (l+1/2)\right]\phantom{mmmm}
\label{Qr}
\ee
contains the terms of order $a^3$ and higher and $\psi(z)$ is the logarithmic derivative of Gamma function \cite{GR}. We emphasize that Eq. (\ref{Tot}) is exact in the parameter $a$. Equation (\ref{Q1}) reflecting the linear one-loop polarization contribution was obtained in \cite{12,as11b,bsaso} and the $a^3$ term in (\ref{Qr}) was  calculated in \cite{bsaso}.
The renormalized induced  charge $q_{ind}$  is negative and odd with respect to  $a$.

In the supercritical range,  we introduce the extension parameter
 $\theta$ instead of $\xi$ \cite{20} accordingly
\bb
\frac{A_R}{\xi A_I} = e^{2i\theta} \left(\frac{2\lambda}{E_0}\right)^{-2i\sigma}
\frac{\nu +a(m+E)/\lambda +is\sigma}{\nu +a(m+E)/\lambda
-is\sigma} \frac{\Gamma(2i\sigma)}{\Gamma(1/2-s/2-aE/\lambda
+i\sigma)} - \frac{\Gamma(-2i\sigma)}{\Gamma(1/2-s/2-aE/\lambda
-i\sigma)}.\label{xithet}
\ee
Here $\pi\geq \theta\geq 0$ and $E_0>0$ is a constant.

Now the Green's function has a discontinuity in the complex plane $E$ and
the quasi-stationary states are on the second (unphysical) sheet with ${\rm Re}\lambda<0$.
In the massless case there emerges the infinite number of quasi-stationary states (resonances) with negative energies determined by complex roots of Eq. ${\rm W}(E, \theta, i\sigma)=0$ \cite{20}.
We calculate the contribution from these resonances in  Eq. (\ref{cursum}) if we integrate  term $j_{super}({\bf r})$ over $E$ on path $S$ along the negative real axis $E$.

Thus, the total induced charge density (\ref{cursum}) can be rewritten as
\bb
j_0(r) = -e\int\limits_{C}\frac{dE}{8\pi^2
i}\sum_{s=\pm 1} \sum^{\infty}_{l=-\infty}\frac{f_I(r, \gamma, E) f_R(r, \gamma, E) + g_I(r, \gamma, E)g_R(r, \gamma, E)}{s{\rm W}(E, \gamma)}-\nonumber \\ -e\int\limits_{S}\frac{dE}{8\pi^2
i}\sum_{l,s: \nu<a} \frac{\xi (f_I^2(r, i\sigma, E) + g_I^2(r, i\sigma, E))}{s{\rm W}(E, i\sigma)}
=q_{ind}(r) +j_{super} (r).\phantom{mmmmmm} \label{Den}
\ee
Here the term $q_{ind}(r)$ is represented by (\ref{Tot}) and the sum over $l$ in $j_{super}$ is taken of  $a^2>(l+s/2)^2$ and the integration path $S$ coincides with the imaginary axis $E$.

 The term $j_{super}$ is convergent, therefore, we can put $m=0$. Summing in $s$, one
obtains
\bb
j_{super}(r) =\frac{e}{4\pi^2 r^2}\sum_{l,\nu<a}
\frac{\nu^2}{\sigma \Gamma(2i\sigma) \Gamma(-2i\sigma)}
\int\limits_{-\infty}^{0} \frac{dE}{E\omega(\sigma)} \Gamma(i\sigma -iaE/|E|)\times \nonumber \\
\times \Gamma(-i\sigma
 -iaE/|E|) W_{iaE/|E| +1/2, i\sigma} (2|E|r)W_{iaE/|E| -1/2, i\sigma}
(2|E|r), \label{supcr1}
\ee
where (here and below) $\nu=l+1/2,\sigma=\sqrt{a^2-(l+1/2)^2}$ and
\bb
\omega(\sigma) = 1- e^{2i\theta} \left(
\frac{2|E|}{E_0}\right)^{-2i\sigma} \frac{\nu+iaE/|E|
+i\sigma}{\nu+iaE/|E| -i\sigma}
\frac{\Gamma(2i\sigma)}{\Gamma(-2i\sigma)}\frac {\Gamma(-i\sigma
 -iaE/|E|)} {\Gamma(i\sigma  -iaE/|E|)}.
\label{omeg1}
\ee
In order to integrate (\ref{supcr1}) over $E$, we substitute $1/r$ for $|E|$ in the factor
$(2|E|/E_0)^{-2i\sigma}\equiv\exp(-2i\sigma \ln (|E|/E_0))$. This can be done because
the integrand (\ref{supcr1}) decreases exponentially as $|E|\gg 1/r$
and strongly oscillate as $|E|$ tends to $0$, hence, the region $|E|\sim 1/r$ mainly contributes
to (\ref{supcr1}).
So,  we need integrate expression
\bb
j_{super}(r) = -\frac{e}{2\pi^2 r^2}\sum_{l, \nu<a}
\frac{\nu^2\Gamma(i\sigma+ia)}{\sigma\omega_-(\sigma) \Gamma(2i\sigma) \Gamma(-2i\sigma)}
\Gamma(-i\sigma +ia) \times \nonumber \\
\times \int\limits_{0}^{\infty} \frac{dE}{E}W_{-ia +1/2,
i\sigma}(2Er)W_{-ia -1/2, i\sigma} (2Er), \phantom{mmmmmmmmmm} \label{supcr2}
\ee
where
\bb
\omega_-(\sigma) = 1- e^{2i\theta +2i\sigma \ln(E_0 r)}
\frac{\nu - ia +i\sigma}{\nu - ia -i\sigma}
\frac{\Gamma(2i\sigma)}{\Gamma(-2i\sigma)}\frac {\Gamma(-i\sigma
+ ia)} {\Gamma(i\sigma  + ia)}. \label{omega1}
\ee
We emphasize that $j_{super}(r)$ is complex quantity, which shows
the instability of neutral vacuum in the supercritical region (see, \cite{grrein}).

Using formula \cite{GR}
\bb
\int\limits_{0}^{\infty} \frac{dE}{E}W_{-ia +1/2,
i\sigma}(2Er)W_{-ia -1/2, i\sigma} (2Er) = \frac{\pi}{\sin(2\pi i\sigma)}\times \phantom{mmmmmmmmmm} \nonumber
\\ \times \left[
\frac{1}{\Gamma(ia +i\sigma) \Gamma(1 +ia
-i\sigma)} -  \frac{1}{\Gamma(ia -i\sigma) \Gamma(1 +ia +i\sigma)} \right]
 \label{F1}
\ee
we  finally obtain
\bb
j_{super}^r(r) = \frac{e}{\pi^2  r^2}\sum_{l,\nu<a}{\rm Re}\frac{\sigma}{\omega_-(\sigma)}. \label{sup4}
\ee
If $1/2<a<3/2$  only the $l=0$ channel is in the supercritical region in which case
\bb
j_{supcr}^r(r) = \frac{e}{\pi^2  r^2}\sigma_0{\rm Re}\frac{2-|A|de^{2i\theta +2i\sigma_0 \ln(E_0 r)+i\psi}}{1-|A|de^{2i\theta +2i\sigma_0 \ln(E_0 r)+i\psi}+|A|^2[(a-\sigma_0)/(a+\sigma_0)]e^{4i\theta +4i\sigma_0 \ln(E_0 r)+2i\psi}}, \label{fsup}
\ee
where
$$
 A=\frac{\Gamma(2i\sigma_0)\Gamma(-i\sigma_0+ia)}{\Gamma(-2i\sigma_0)\Gamma(i\sigma_0+ia)},\quad
 d=2\frac{a-\sigma_0}{a},\quad \sigma_0=\sqrt{a^2-1/4},
$$
$$
\psi \equiv {\rm Arg} A = -\pi-2{\cal C}\sigma_0 +\sum\limits_{n=1}^{\infty}\left(\frac{2\sigma_0}{n}-2\arctan\frac{2\sigma_0}{n}
+\arctan\frac{2n\sigma_0}{n^2+1/4}\right).
$$
Here ${\cal C}=0.57721$ is Euler's constant.

%We see that the induced vacuum polarization for noninteracting massless fermions
%has a power law form ($\sim c/r^2$)  whose coefficient is log-periodic functions with respect
%to the distance from the origin.

For small $\sigma_0\ll 1$, Eq. (\ref{fsup}) takes the simplest form
\bb
j_{supcr}^r(r) = \frac{e\sigma_0}{\pi^2  r^2}. \label{1fsup}
\ee
This expression was obtained in \cite{as11b} by means of the exact phase-shifts analysis.

%The induced charge density $j_{super}^r(r)$ (\ref{1fsup})  at $\sigma_0\ll 1$ does not contain at
%all the self-adjoint extension parameter $\theta$. From the physical point of view,
%when  the Coulomb center charge is suddenly increased from subcritical
%to supercritical values,  the transition will occur from the subcritical region to the supercritical
%one,  and then a small change in $a$ such that $a>a_{cr}$ leads to a sudden change
%in the character of a physical phenomenon due to emerging of the resonances with negative energies.

The emerging  resonances  may significantly  shield a  Coulomb impurity in graphene.
Indeed, an electron at distance $r$ from the Coulomb center feels
the effective charge that is the charge impurity minus
the induced screening charge $q(r)$ within the annulus $r_0, r,\quad r_0<r$.
 For small $\sigma_0$ $q(r)$ can be found by integrating Eq. (\ref{1fsup})
\bb
q(r)=-2\frac{e_0 \sigma_0}{\pi}\ln\frac{r}{r_0}.
\label{charge0}
\ee

 We can rewrite Eq. (\ref{charge0}) for the effective coupling
$g\equiv a_{eff}$  like the differential equation of the renormalization group (see \cite{as11b,review}):
\bb
\frac{dg}{d\ln(r/r_0)}=-2\frac{e_0^2 \sigma_0}{\pi}.
\label{renorm}
\ee
We see that the effective coupling $g$  will tend to the critical value $g_{cr}=1/2$  at
finite distances
$$
r=r_0e^{-(2\pi/e_0^2)\ln[2g+\sqrt{4g^2-1}]}
$$
from the Coulomb impurity.
The renormalization group treatment is applicable when the right of equation (\ref{renorm}) is small.
%Therefore for $\sigma_0\ll 1$,  the vacuum of planar electrons with Coulomb impurity
%self-consistently rearranges itself so that electrons at distances $r>r_0$ never feel
%a supercritical  effective charge irrespective of the charge impurity \cite{as11b,review}.

 It is essential that the number (critical charge), energy spectrum as well as lifetime
of emerging resonances depends weakly upon $\theta$ at $\sigma_0\ll 1$.
Since the induced charge density does not depend upon the parameter $\theta$ that can be related to the radius $R$ of a supercritical impurity, one can conclude that the impurity size does not affect the induced charge density near the transition point ($\gamma=i\sigma, \sigma\ll 1$) at large distances $r\gg R$. We emphasize that it is not the case for massive fermions.

For large $a\gg 1, \sigma\approx a-l^2/2a$, the induced charge density can approximately
be represented as
\bb
{\rm Re}j_{super}(r) = \frac{e}{\pi^2 r^2}\sum_{l<a}\sqrt{a^2-l^2},\quad {\rm Im}j_{super}(r) =0. \label{im0sup}
\ee

\section{Vacuum polarization of planar charged massive fermions}

We now briefly  address to the vacuum polarization induced by the Coulomb potential
in massive case. If the  Coulomb center charge is subcritical the massive case has a well defined infinite spectrum of bound solutions situated on the physical sheet, which for $\gamma\geq 1/2, a<1/2, \xi=0$ is \cite{khho}
 \bb
 E_{k,l} = m\frac{k +  \sqrt{\nu^2-a^2}}
 {\sqrt{[k + \sqrt{\nu^2-a^2}]^2+a^2}}, \quad \nu =l+1/2;\quad k, l= 0, 1, 2 \ldots ,
\label{spectrumw}
\ee
We see that all the energy levels are doubly degenerate with respect to  $s$.
It can be easily shown  that
the spectrum accumulates at the point $E = m$, and its asymptotic form as
$n=k+l\to \infty$ is given by the nonrelativistic formula
$\epsilon_n=m-E_n=ma^2/n^2$. The problem of finding the spectra of self-adjoint extensions of the radial Hamiltonian in the Coulomb and Aaronov-Bohm potentials
in 2+1 dimensions was solved in \cite{khlee0} where, in particular, it was shown that
the spectrum accumulates at the point $E = m$ and is described by the same asymptotic
formula (without AB potential), independent of $\xi$, i.e. $\epsilon_n=m-E_{n,\xi}=ma^2/n^2$.

In the massive case the vacuum polarization of planar charged fermions manifests itself by modifying the Coulomb potential. Therefore, it is rewarding to calculate the polarization corrections to the Coulomb potential. As applied to the vacuum polarization we shall assume that none of the bound levels are occupied. If $a\ll 1$ we can estimate these polarization corrections in the first order in $a$.
For three spatial dimensions, the potential taking into account the polarization corrections of the first order in $a$ to the Coulomb potential is the Uehling-Serber potential.
In terms of perturbation theory, these corrections correspond to the polarization operator
in the lowest order in interaction. Performing the integrations and summation in Eq. (\ref{Imp}) with taking  only the linear in $a$ terms into account,  for the renormalized induced
Coulomb center charge, we obtain
\bb
q_m(|{\bf p}|)=-\frac{a}{e_0}\frac{\Pi(-{\bf p}^2)}{|{\bf p}|},
\label{po}
\ee
where, as it should be,
$$
\Pi(-{\bf p}^2)=\frac{e_0^2}{8\pi}\left(\frac{4m^2-{\bf p}^2}{\sqrt{{\bf p}^2}}\arctan\sqrt{\frac{{\bf p}^2}{4m^2}}-2m\right)
$$
is the polarization operator in the first order of perturbation theory.
After some transformations the induced charge distribution $q_m(r)\equiv a_{eff}^m/e_0$
(here $a_{eff}^m$ is the effective coupling) takes the form in the coordinate space:
\bb
q_m(r)=-e_0a\int\limits_{1}^{\infty}\frac{dx}{x^3\sqrt{x^2-1}}e^{-2mrx}.
\label{q1c}
\ee
The integral is calculated in limits $mr\ll 1$ and $mr\gg 1$ and as a result we find
\bb
q_m(r)\approx -   e_0a\left[\frac{\pi}{4}-Cmr\right], \quad mr\ll 1, 1\gg C\gg mr,
\label{smalmr}
\ee
where the first term on the right of Eq. (\ref{smalmr}) was already calculated (see Eq. (\ref{Q1})), and
\bb
q_m(r)\approx -e_0a\sqrt{\frac{4\pi}{mr}}e^{-2mr},  \quad mr\gg 1.
\label{bigmr}
\ee
We see that even at small distances from the Coulomb center, the finite mass contribution to the
induced vacuum charge is small and insignificantly distorts the Coulomb potential only at distances of the Compton length $r\sim 1/m$. The  induced charge   has a screening sign.

In the supercritical regime the finite mass contribution to the vacuum polarization easier to
estimate,  at least when $\sigma_0\equiv \sqrt{a^2-1/4}\ll 1$.
Indeed, if the Coulomb potential charge is suddenly increased from subcritical
to supercritical values then the only  lowest energy level  dives into the
negative energy continuum and becomes a resonance with ``complex energy''$E=|E|e^{i\tau}$.
 There appears the pole on the unphysical sheet $\tau>\pi$, counted now as a ``hole'' state.
Using results of Ref. \cite{khlee0}, one can show  the energy of dived state
${\rm Re}E=-(m+\epsilon), \epsilon\to +0$,  is determined by the following transcendental equation
\bb
 \arg\Gamma(2i\sigma_0)-\sigma_0{\rm Re}\psi(-ix)-(\sigma_0/2)\ln(8\epsilon/m)+\arctan[\sigma_0(1-2a^2\epsilon/m)]=-\theta,
\label{bound-men}
\ee
where $x=\sqrt{ma^2/2\epsilon}$.
This resonance is spread out over an energy range of order
$\Gamma_g \sim me^{-\sqrt{2m\pi a^2/\epsilon}}$ and strongly distort
around the Coulomb center. The resonance is sharply defined state with diverging
lifetime $(\Gamma_g)^{-1}\sim e^{\sqrt{2m\pi a^2/\epsilon}}/m$.
Thus, the resonance is practically a bound state.

 The diving point for the energy level defines and depends upon the
parameter $\theta$. This diving of bound levels entails a complete restructuring
of the quantum electrodynamics vacuum in the supercritical Coulomb field \cite{003,grrein}.
As a result, the QED vacuum  acquires the charge, thus leading to the concept of a charged
vacuum in supercritical fields due to the real vacuum polarization \cite{003,grrein}. As was shown
in \cite{grrein} the contribution  to the Green's function from the only pole on the second sheet contains the only term associated with the former lowest bound state:
\bb
G_r({\bf r}, {\bf r'}; E)=i\frac{\Gamma_g\Theta(-m-E)}{(E-E_0)^2+\Gamma_g^2/4}\psi^{cr}_0({\bf r})[\psi^{cr}_0({\bf r'})]^{\dagger},
\label{grsup}
\ee
where $\Theta(z)$ is the step function and $\psi^{cr}_0({\bf r})$ is the ground state of the Dirac Hamiltonian at $a=a_{cr}$ (the critical state) with energy $E_0$ within the gap $-m\leq E_0<m$ but close to $-m$. The  critical  charge $a_{cr}$ is defined as the condition for the appearance of the imaginary part of ``the energy''.  It is important that the Green function of the type (\ref{grsup})
 eliminates the lack of stability of  neutral vacuum for $a>a_{cr}$ (see, \cite{grrein}).
Then, the real vacuum polarization charge density can be determined by
\bb
 j_0^{real}({\bf r})\equiv -\frac{e_0}{2}\int\limits_{R}\frac{dE}{2\pi i}{\rm tr}G({\bf r}, {\bf r'}; E)|_{{\bf r}={\bf r'}}\gamma_0,
\label{cur12}
\ee
where the path $R$ surrounds the  singularity on the unphysical sheet.
Integrating (\ref{cur12}) we obtain $j_0^{real}({\bf r})=-e_0|\psi^{cr}_0({\bf r})|^2$.

We see that the space density of the real vacuum polarization is real quantity
and approximately described with the modulus squared of the fermion wave function
in the critical state:
$$
j_0^{real}(r) \sim - e_0m^2[2(\ln mr)^2-2(\ln mr)/a_{cr} +1/a_{cr}^2], \quad mr\ll 1
$$
and
$$
j_0^{real}(r)\sim -e_0me^{-2\sqrt{r/l}}/r, \quad l=1/\sqrt{2m\epsilon_0}, mr\gg 1,
$$
where $\epsilon_0$ depends upon $a_{cr}$ and the extension parameter $\theta$.

The total induced charge density in massive case with taking into account the real vacuum polarization  (\ref{cur12}) can be estimated as  the sum: $q_m(r)m^2+j_0^{real}$.

\section{Conclusion}

In this paper we obtain an expression for the density of a planar induced vacuum charge
in a strong Coulomb potential in coordinate space. The treatment is based
on a self-adjoint extension approach.  For the first time  we express
the density of a planar induced charge in the vacuum via the exact Green function,
constructed from solutions of the self-adjoint two-dimensional Dirac Hamiltonians
with a strong Coulomb potential. Induced vacuum  charge density is calculated
and analyzed at the subcritical and supercritical Coulomb potentials for massless
and massive fermions.  The behavior of the obtained vacuum charge density is investigated
at long and short distances from the Coulomb center.

In  the subcritical range for $m=0$, the induced vacuum charge $q_{ind}$ is obtained
as an exact odd function of the Coulomb coupling $a$.

  For the first time we express the induced vacuum charge in the supercritical Coulomb potential via
the exact Green function, which has the singularities (on the nonphysical sheet of the Riemann surface) on the negative energy axis related to the creation of infinitely many quasi-stationary states. We discuss screening of the supercritical Coulomb impurity in graphene.

In the massive case, we argue that the contribution into the induced vacuum charge coming from terms containing the mass $m$ is small  compared with massless terms and insignificantly distorts
the Coulomb potential only at distances of order of the Compton length $1/m$.
The  induced vacuum charge has a  screening sign.
As is known the quantum electrodynamic vacuum becomes unstable when the Coulomb center charge
is increased from subcritical to supercritical values. In the massive case,
when the Coulomb center charge becomes supercritical
then the lowest state turn into resonance with a diverging lifetime,
which can be described as a quasi-stationary state with ``complex energy'';
the quantum electrodynamics vacuum acquires the charge due to the so-called real
vacuum polarization. An expression for the real vacuum polarization charge density is obtained
in a supercritical Coulomb potential.

We  briefly discuss  the vector Coulomb problem in the presence of scalar Coulomb
potential and argue that the quantum electrodynamics vacuum  in the vector Coulomb potential
is stabilized in the presence of a scalar potential.

\vspace{0.5cm}

{\bf Appendix: Charge renormalization}

\vspace{0.3cm}
We represent  $q_{ind}$ in the form (\ref{Tot}) (see, also \cite{ms0}, where the induced
vacuum charge in a subcritical Coulomb field was calculated in
the conventional three-dimensional massive quantum electrodynamics).

At first we calculate $q_1(e_0a)$. This term is obtained from (\ref{Imp}) by the substitution
$g(y,t)\to g_1 (y,t)$, where $g_1 (y,t)$ is
\begin{eqnarray}
g_1 (y,t) = 4a \left[ \frac{ytx^2}{1+x^2} I^{\prime}_{2\nu}(y) - \coth t I_{2\nu}(y)\right]. \label{f1}
\end{eqnarray}
Then, taking into account
\begin{eqnarray}
\int\limits^{\infty}_{0} dy I_{2\nu} (y) e^{-y \cosh t} =
e^{-2 \nu t}/\sinh t, \label{Fom1}
\end{eqnarray}
we see that the right of equation (\ref{Imp}) diverges when $z\rightarrow 0$:
\begin{eqnarray}
-\frac{4ea}{\pi} \sum^{\infty}_{l=0} \int^{\infty}_{0} dt \coth t
 e^{-2 \nu t}.\label{Div}
\end{eqnarray}
Diverging term should be subtracted from the integrand (\ref{Imp}) with
 $g_1 (y,t)$ and we obtain:
\begin{eqnarray}
\rho_r^1(z) = \frac{2e}{\pi} \sum_{l=0}^{\infty}
\int\limits^{\infty}_{0} dt \left( \int\limits^{\infty}_{0} dy
\int\limits^{\infty}_{0} dx \frac{\sinh t}{2\sqrt{1+x^2}} e^{-y \cosh t}
J_{0} (z y \sinh t/2\sqrt{1+x^2}) g_1 (y,t) + 2a \coth t
 e^{-2 \nu t} \right). \label{Reg}
\end{eqnarray}

Taking account of that when  $m\to 0$
\begin{eqnarray}
\int\limits^{\infty}_{0} dt \int\limits^{\infty}_{0} dy
\frac{\sinh t}{2\sqrt{1+x^2}} e^{-y \cosh t} g_1(y,t) = 0, \label{EQ}
\end{eqnarray}
rewrite (\ref{Reg}) as:
\begin{eqnarray}
q_1(e_0a) = \frac{2e}{\pi} \sum_{l=0}^{\infty} \int\limits^{\infty}_{0}
dt \left( \int\limits^{\infty}_{0} dy \int\limits^{\infty}_{0} dx
\frac{\sinh t}{2x} e^{-y \cosh t} g_1 (y,t) [J_{0} (y \sinh t/x)
-J_{0} (1/x)] +
\right. \nonumber \\
\left. + 2a \coth t e^{-2 \nu t}  \right). \label{Imp1}
\end{eqnarray}
Integrating over $x$, we obtain:
\begin{eqnarray}
q_1(e_0a) = \frac{2ea}{\pi} \sum_{l=0}^{\infty} \int\limits^{\infty}_{0}
dt \left( \int\limits^{\infty}_{0} dy \sinh t \ln (1/y \sinh t)
e^{-y \cosh t}
 \left[ yt I^{\prime}_{2\nu} (y) - \coth t I_{2\nu} (y)\right]
+ a \coth t  e^{-2 \nu t}  \right)= \nonumber \\
= \frac{4ea}{\pi} \sum_{l=0}^{\infty} \int\limits^{\infty}_{0} dt
\left( \int\limits^{\infty}_{0} dy \ln (1/y \sinh t) \left[ t
\sinh t \frac{d}{dy} [y e^{-y \cosh t} I_{2\nu} (y)] -\right.\right. \phantom{mmmmm}\nonumber \\
\left.\left. -\frac{d}{dt} ( t \cosh t  e^{-y \cosh t})
I_{2\nu} (y)\right] + a \coth t
 e^{-2 \nu t}  \right). \phantom{mmmmmmmmmm}\label{Imp2}
\end{eqnarray}
Integrating this expression over $y$ and then over $t$, we obtain (\ref{Q1}).

We now renormalize the terms of order $a^3$ and higher.
At first, we subtract from the integrand (\ref{Imp}) the terms linear in
$a$, given with function $g_1(y, t)$, and represent the result as:
\bb \rho_r^h (z)=\frac{2e}{\pi}
\sum_{l=0}^{\infty} \left[ \int\limits^{\infty}_{0} dx
\int\limits^{\infty}_{0} dt \int\limits^{\infty}_{0} dy
\frac{\sinh t}{2\sqrt{1+x^2}} e^{-y \cosh t} J_{0} (zy \sinh t/2\sqrt{1+x^2})
[g(y,t)-g_1 (y,t)] - f(l) \right], \label{reg2} \ee where \bb
f(l) = \lim_{z\to 0} \int\limits^{\infty}_{0} dx
\int\limits^{\infty}_{0} dt \int\limits^{\infty}_{0} dy
\frac{\sinh t}{2\sqrt{1+x^2}} e^{-y \cosh t} J_{0} (z y \sinh t/2\sqrt{1+x^2})
[g(y,t)-g_1 (y,t)] \label{omegal}  \ee
involves the asymptotic form of the terms of order $a^3$ and higher
 as $z\to 0$.
Let us calculate $f(l)$. Applying (\ref{Fom1}) and
$J_{0} (z y \sinh t/2\sqrt{1+x^2})|_{z\to 0}\rightarrow 1$, we integrate (\ref{omegal})
in $y$ and rewrite expression obtained in the form: \bb f(l) =
2\int\limits^{\infty}_{0} \frac{dx}{\sqrt{1+x^2}} \int\limits^{\infty}_{0} dt
\left[ -\frac{x}{2\sqrt{1+x^2}} \sin(k't) \frac{d}{dt} (e^{-2\gamma t}\coth t) - a \cos (k't) \coth t
e^{-2\gamma t} + \right. \nonumber \\
 \left. + a t \frac{x^2}{1+x^2} \frac{d}{dt} (e^{-2\nu t} \coth
 t)+ a \coth t e^{-2\nu t} \right].
\label{omegal1} \ee

Integration by parts the terms with derivative in $t$ and then
integration of obtained expression over $x$ gives:
\bb f(l) = 2\int\limits^{\infty}_{0} dt \left[ a e^{-2\nu t} - \frac{\sin (2at)}{2t} e^{-2\gamma t}
 \right] \coth t. \label{omegal2} \ee
Having been differentiated  with respect to $a$ the obtained expression was integrated over $t$
with using formula: \bb \int\limits^{\infty}_{0}  dt e^{-2\nu t }
\left( \frac{1}{t} - \coth t \right) = \psi (\nu) - \ln (\nu)
+\frac{1}{2\nu}. \label{FomImp} \ee Then, integrating it over $a$ with
taking account of the obvious boundary condition
($f(l)=0$ at $a=0$), we obtain the final expression: \bb f(l) = - 2{\rm Im} \left[ \ln (\Gamma
(\gamma- ia) + \ln (\gamma - ia) +  ia\psi(l+1/2) +
\frac{ia}{2(l+1/2)} \right]. \label{omegaldef} \ee

Now we consider (\ref{reg2}) at the limit $m\to 0$.
Taking into account that formula (\ref{EQ}) is also valid for
$g (y, t)$ at  this limit, we rewrite Eq. (\ref{reg2}) as follows:
\bb q_r(e_0a)=\frac{e}{\pi} \sum_{l=0}^{\infty} \left[
\int\limits^{\infty}_{0} dx \int\limits^{\infty}_{0} dt
\int\limits^{\infty}_{0} dy \frac{\sinh t}{x} e^{-y \cosh t}
[J_{0} (y \sinh t/x) - J_0 (1/x)] [g(y,t)-g_1 (y,t)] - f(l)
\right].
 \label{QRInt} \ee
At first we integrate this expression over $x$
and then  over $y$ and $t$ with using Eqs. (\ref{FomImp}),  (\ref{Fom1}).
As a result, we obtain: \bb q_r(e_0a) =
-\frac{2e}{\pi} \sum^{\infty}_{l=0} [2{\rm Im}[
(\gamma -ia) \psi (\gamma -ia) + ia\psi(l+1/2) + ia (l+1/2)\psi^{\prime}(l+1/2)] - f(l)]. \label{QRdef} \ee

Substituting  $f(l)$ from (\ref{omegaldef}), we obtain (\ref{Qr}).

\end{document}